# Stimulated Bremsstrahlung in Electron–Nucleus Scattering in a Multifrequency Electromagnetic Field

**S. P. Roshchupkin and A. I. Voroshilo**
*Sumy State University, Sumy, 244007 Ukraine*

**Abstract**—We theoretically investigate stimulated bremsstrahlung emission and absorption (SBEA) in the scattering of a relativistic electron by a nucleus in a multifrequency field of circularly polarized plane electromagnetic waves propagating in the same direction. The general relativistic expression for the SBEA probability is derived. It is demonstrated that the probability of multiphoton SBEA involving an arbitrary number of waves $N$ is governed by the function $I_{n_1 \dots n_N}$ defined by (17) and (18), which depends on Bunkin–Fedorov quantum parameters $\gamma_k$ (20) and quantum interference parameters $\alpha_{kj}$ (21). We separate two substantially different kinematic ranges in electron scattering: the Bunkin–Fedorov range, where quantum parameters $\gamma_k$ serve as multiphoton parameters, and the interference range, where the interference of all the waves becomes clearly pronounced, and quantum parameters $\alpha_{kj}$ play the role of multiphoton parameters. In the interference range, we reveal a correlation between the numbers of photons in all the waves. Due to this correlation, the total number of photons emitted or absorbed by an electron (with allowance for the polarizations of waves) is equal to zero. We analyze the cases of nonrelativistic, relativistic, and ultrarelativistic electron energies. It is demonstrated that the probability of multiphoton SBEA for an electron scattered in the interference range may considerably exceed the corresponding probability in the Bunkin–Fedorov range.

## 1. INTRODUCTION

The process of stimulated bremsstrahlung emission and absorption (SBEA) accompanying electron–nucleus scattering in the field of a plane monochromatic electromagnetic wave has been thoroughly investigated (e.g., see [1–6]). Studies [7–9] provide a detailed analysis of this process in the field of two elliptically polarized light waves with arbitrary intensities and frequencies propagating in the same direction. It was demonstrated that, depending on intensities, frequencies, and polarizations of electromagnetic waves, electron–nucleus scattering may occur in various kinematic ranges and can be characterized by various quantum multiphoton parameters. These studies revealed an interference effect, which is manifested in circularly polarized waves in the case of substantially asymmetric scattering and which is accompanied by correlated emission and absorption of equal numbers of photons from both waves.

In this paper, we extend the results of previous papers to the case of an arbitrary number of plane monochromatic waves propagating in the same direction. We consider the case of circularly polarized waves, when the interference of waves becomes clearly pronounced. We will demonstrate that the increase in the number of waves gives rise to a nontrivial effect of electron emission and absorption at combination frequencies $|\Omega_{kN}|$ (21), which is associated with a strong correlation between the numbers of photons in all the waves.

In Section 2, we derive the amplitude of electron–nucleus SBEA in the field of $N$ circularly polarized electromagnetic waves in the general relativistic case. It is demonstrated that multiphoton processes of emission and absorption of $n_1, n_2, \dots, n_N$ photons from the 1st, 2nd, …, $N$th waves ($n_1, n_2, \dots, n_N$ are negative and positive integer numbers) are governed by the function $I_{n_1 \dots n_N}$ defined by (17) and (18). This function depends on $N$ Bunkin–Fedorov quantum parameters $\gamma_j$ (20) and on

$$N_1 = N(N-1)/2 \tag{1}$$

quantum interference parameters $\alpha_{kj}$ (21). Under conditions when interference parameters are small ($\alpha_{kj} \ll 1$), $n_1, n_2, \dots, n_N$ photons are independently emitted and absorbed in each wave [the functions $I_{n_1 \dots n_N}$ can be represented as products of independent Bessel functions for each wave (22)]. Provided that $\alpha_{kj} \gtrsim 1$, emission and absorption of $n_1, n_2, \dots, n_N$ photons of electromagnetic waves are generally accompanied by all possible virtual processes involving emission and absorption of photons at combination frequencies (21). We will write the SBEA amplitude in a compact form in terms of operators $b_j^{(\pm)}$ (15) and functions $I_{n_1 \dots n_N}$.

In Section 3, we derive the probability of electron–nucleus SBEA in the field of $N$ waves in the general relativistic case. The SBEA probability is represented as a sum of partial probabilities to emit and absorb $n_1, n_2, \dots, n_N$ photons of the waves. We demonstrate that it is possible to integrate the relevant partial probability in the quasienergy of the electron in the final state.

In Section 4, we analyze electron–nucleus SBEA in the Bunkin–Fedorov range (30), where an electron experiences scattering within a rather broad kinematic region (excluding only the case when an electron is scattered by a preset angle in the plane of the incident electron momentum and the wave vector). In this case, quantum parameters $\gamma_k$ (20) serve as multiphoton parameters. It is demonstrated that, for moderately strong fields [$\xi_k \ll 1$, $k = 1, \dots, N$ (31)], the cross section of multiphoton electron–nucleus SBEA in the general relativistic case can be factorized as a product of the probability to emit or to absorb a photon from the relevant wave, which is determined by the modulus of functions $I_{n_1 \dots n_N}$ squared, by the cross section of elastic electron–nucleus scattering (the Mott cross section). In the limiting case of nonrelativistic electron velocities, we find the condition of the applicability of dipole approximation (38). We also thoroughly investigate the range of strong fields ($\xi_k \sim 1$) for nonrelativistic electrons.

In Section 5, we consider electron–nucleus SBEA in the interference range (50), where an electron is scattered by a preset angle in the plane of the initial electron momentum and the wave vector. In this regime, the interference of waves becomes clearly pronounced, and multiphoton parameters are defined as quantum interference parameters $\alpha_{kj}$ (21) (Bunkin–Fedorov parameters are small for arbitrary wave intensities). It is demonstrated that, in the interference range, we can introduce ($N$ – 1) sources of photons at combination frequencies $|\Omega_{kN}|$ ($k = 1, 2, \dots, N - 1$) (21). Then, functions $I_{n_1 \dots n_N}$ (17) can be considerably simplified, and multiphoton processes are governed by functions $J_{l_1 \dots l_{N-1}}$ (57), which depend only on interference quantum parameters. It is demonstrated that, within the range of light fields attainable with modern sources of coherent radiation, the probability of multiphoton electron–nucleus SBEA for nonrelativistic, relativistic, and ultrarelativistic energies may considerably exceed the probability of scattering in the Bunkin–Fedorov range. The cross sections of these processes are determined.

## 2. THE SCATTERING AMPLITUDE

Let us choose the four-potential of the external field in the form of a sum of $N$ circularly polarized electromagnetic waves propagating along the $z$-axis:

$$A = \sum_{j=1}^{N} A_j(\varphi_j), \qquad (2)$$

where

$$A_j(\varphi_j) = \frac{F_j}{\omega_j}(e_{jx}\cos\varphi_j + \delta_j e_{jy}\sin\varphi_j).^1 \qquad (3)$$

Here, $\delta_j = \pm 1$; $e_{jx} = (0, \mathbf{e}_{jx})$, and $e_{jy} = (0, \mathbf{e}_{jy})$ are the polarization four-vectors; $F_j$ and $\omega_j$ are the strength and the frequency of the $j$th wave, respectively; and the argument $\varphi_j$ is written as

$$\varphi_j = \omega_j(t - z), \quad j = 1, 2, \dots, N. \qquad (4)$$

In the first Born approximation in the interaction of the electron with the field of the nucleus $Ze$, the amplitude of electron scattering in the field of $N$ waves (2) is described by the following expression [2, 7, 8, 10]:

$$S_{fi} = -ie\int d^4x \bar{\psi}_f(x|A)\left[\tilde{\gamma}_0 \frac{Ze}{|\mathbf{x}|}\right]\psi_i(x|A), \qquad (5)$$

where $\psi_i$ and $\bar{\psi}_f$ are the wave functions (Volkov functions) of the electron in the field of $N$ waves before and after scattering, respectively, and $\tilde{\gamma}_\mu$ ($\mu = 0, 1, 2, 3$) are the Dirac matrices. Expanding Volkov functions in Fourier series and evaluating the relevant integrals, we can represent the sought-for amplitude in the form

$$S_{fi} = \left[\prod_{j=1}^{N} \sum_{n_j=-\infty}^{\infty}\right] S_{fi}^{(n_1 \dots n_N)}, \qquad (6)$$

where the partial amplitude corresponding to emission ($n_1 > 0$, $n_2 > 0$, …, $n_N > 0$) and absorption ($n_1 < 0$, $n_2 < 0$, …, $n_N < 0$) of $n_1, n_2, \dots, n_N$ photons from the 1st, 2nd, …, $N$th waves, respectively, is given by

$$S_{fi}^{(n_1 \dots n_N)} = -i\frac{4\pi^2 Ze^2}{\sqrt{\tilde{E}_i \tilde{E}_f}} \times \exp(i\phi_{fi})[\bar{u}_f M_{fi}^{(n_1 \dots n_N)} u_i]\frac{\delta(q_0)}{\mathbf{q}^2}. \qquad (7)$$

Here, $\phi_{fi}$ is the phase independent of summation indices; $u_i$ and $\bar{u}_f$ are the Dirac bispinors; and $q_0$ and $\mathbf{q}$ are the components of the transferred four-momentum $q = (q_0, \mathbf{q})$, which can be represented as

$$q = \tilde{p}_f - \tilde{p}_i + \left(\sum_{j=1}^{N} n_j\omega_j\right)n, \qquad (8)$$

where $n = (1, \mathbf{n})$ ($\mathbf{n}$ is the unit vector along the direction of propagation of all $N$ waves; we should also recall that $\omega_j n$ is a four-vector) and $\tilde{p}_{i,f} = (\tilde{E}_{i,f}, \tilde{\mathbf{p}}_{i,f})$ are the four-quasimomenta of the electron before and after scattering [10],

$$\tilde{p}_{i,f} = p_{i,f} + \frac{m^2}{2\kappa_{i,f}}\left(\sum_{j=1}^{N}\eta_j^2\right)n, \quad \kappa_{i,f} = E_{i,f} - \mathbf{n}\mathbf{p}_{i,f}. \qquad (9)$$

Here, $p_{i,f} = (E_{i,f}, \mathbf{p}_{i,f})$ are the four-momenta of the electron before and after scattering and $\eta_j$ is the classical relativistic-invariant parameter of the $j$th wave,

$$\eta_j = eF_j/m\omega_j, \quad j = 1, 2, \dots, N. \qquad (10)$$

[1] We use the relativistic system of units: $\hbar = c = 1$.

Evaluating the relevant sums, we can represent the amplitude $M_{fi}^{(n_1 \dots n_N)}$ in (7) in the following form:

$$M_{fi}^{(n_1 \dots n_N)} = \tilde{\gamma}^0 I_{n_1 \dots n_N} + \frac{m^2}{2\kappa_i \kappa_f} B_{n_1 \dots n_N} \hat{n}$$
$$+ \frac{m}{2\kappa_i} \hat{D}_{n_1 \dots n_N} + \frac{1}{4}\left(\frac{m}{\kappa_f} - \frac{m}{\kappa_i}\right) \hat{D}_{n_1 \dots n_N} \hat{n} \tilde{\gamma}^0. \quad (11)$$

Here, as usually, the caps denote scalar products of four-vectors with Dirac matrices ($\hat{D} = D_\mu \tilde{\gamma}^\mu$), and the four-vector $D_{n_1 \dots n_N}$ and functions $B_{n_1 \dots n_N}$ are written as

$$D_{n_1 \dots n_N} = \sum_{j=1}^{N} \eta_j [e_j b_j^{(-)} + e_j^* b_j^{(+)}] I_{n_1 \dots n_N},$$
$$e_j = e_{jx} - i\delta_j e_{jy}, \quad (12)$$

and

$$B_{n_1 \dots n_N} = \left(\sum_{j=1}^{N} \eta_j^2\right) I_{n_1 \dots n_N}$$
$$+ \sum_{k=1}^{N-1} \left\{ \sum_{j=k+1}^{N} \eta_k \eta_j [\exp(-i\delta_k \Delta_{kj}) b_k^{(-)} b_j^{(\delta_k \delta_j)} \right. \quad (13)$$
$$\left. + \exp(i\delta_k \Delta_{kj}) b_k^{(+)} b_j^{(-\delta_k \delta_j)}] \right\} I_{n_1 \dots n_N},$$

$$\Delta_{kj} = \angle(\mathbf{e}_{kx}, \mathbf{e}_{jx}). \quad (14)$$

To make the notation more convenient, we introduced operators $b_j^{(\pm)}$ in (12) and (13). These operators act on the index $n_j$ of functions $I_{n_1 \dots n_j \dots n_N}$, increasing or decreasing these numbers by one:

$$b_j^{(\pm)} I_{n_1 \dots n_j \dots n_N} = I_{n_1 \dots (n_j \pm 1) \dots n_N}. \quad (15)$$

Therefore, each term in (12) involves functions $I_{n_1 \dots (n_j \pm 1) \dots n_N}$, and each term in (13) involves functions $I_{\dots (n_k \pm 1) \dots (n_j \pm 1) \dots}$. These functions depend on

$$N_2 = N(1 + N)/2 \quad (16)$$

quantum parameters and can be represented as $N_1$ (1) sums of products involving $N_2$ integer-order Bessel functions $J_{n_k}$:

$$I_{n_1 \dots n_N}(\gamma_1, \dots, \gamma_N; \alpha_{12}, \dots, \alpha_{kj}, \dots, \alpha_{N-1,N})$$
$$= \exp\left\{-i\sum_{j=1}^{N} \delta_j \chi_j n_j\right\} \left[\prod_{k=1}^{N-1} \prod_{j=k+1}^{N} \sum_{r_{kj}=-\infty}^{\infty}\right] C^{(r_{12} \dots r_{N-1,N})}, \quad (17)$$

$$C^{(r_{12} \dots r_{N-1,N})} = \left\{\prod_{k=1}^{N} J_{\lambda_k}(\gamma_k)\right\} \left[\prod_{k=1}^{N-1} \prod_{j=k+1}^{N} J_{r_{kj}}(\alpha_{kj})\right], \quad (18)$$

$$\lambda_k = \delta_k \left[\sum_{j=k+1}^{N} r_{kj} - \sum_{j=1}^{k-1} r_{jk}\right] - n_k;$$
$$\chi_k = \angle(\mathbf{e}_{kx}, \mathbf{g}_{fi}''). \quad (19)$$

Here, $\gamma_k$ are the well-known Bunkin–Fedorov quantum multiphoton parameters for each wave,

$$\gamma_k = \eta_k \frac{m}{\omega_k} |\mathbf{g}_{fi}| \sin\psi, \quad \psi = \angle(\mathbf{n}, \mathbf{g}_{fi}),$$
$$\mathbf{g}_{fi} = \frac{\mathbf{p}_f}{\kappa_f} - \frac{\mathbf{p}_i}{\kappa_i}, \quad k = 1, 2, \dots, N, \quad (20)$$

and $\alpha_{kj}$ are the quantum interference parameters [the number of such parameters is equal to $N_1$ (1)],

$$\alpha_{kj} = \eta_k \eta_j \frac{m^2}{\Omega_{kj}} \left(\frac{1}{\kappa_f} - \frac{1}{\kappa_i}\right); \quad \Omega_{kj} = \delta_k \omega_k - \delta_j \omega_j. \quad (21)$$

In (19), $\mathbf{g}_{fi}'' = \mathbf{g}_{fi} \sin\psi$ is the projection of the vector $\mathbf{g}_{fi}$ on the polarization plane of the waves. Note that Bunkin–Fedorov quantum parameters (20) and interference parameters (21) can be represented in the relativistic-invariant form if we take into account that $\omega_j \kappa_{i,f}$ is invariant. As can be seen from (21), if two waves have identical polarizations ($\delta_k \delta_j = 1$; i.e., the strength vectors of these fields rotate in the same direction), then interference parameters involve the difference of the relevant frequencies. If these waves have different polarizations ($\delta_k \delta_j = -1$; i.e., the strength vectors of these fields rotate in opposite direction), then interference parameters involve the sum of the relevant frequencies. Note also that, in the limiting case of nonrelativistic electron velocities, interference parameters take into account the nondipole character of the interaction of an electron with electric wave fields (in the dipole approximation, we have $\kappa_f = \kappa_i = m$, and, consequently, $\alpha_{kj} = 0$).

As can be seen from expressions (7), (11)–(13), and (17), to describe the effect of many waves, we should not only formally introduce the terms responsible for these waves but also take into account interference terms (the terms proportional to the products $\eta_k \eta_j$ of wave intensities). Generally, the functions $I_{n_1 \dots n_j \dots n_N}$, which determine the probabilities of multiphoton emission and absorption processes, cannot be represented as products of independent Bessel functions $J_{n_1}(\gamma_1) \dots J_{n_N}(\gamma_N)$. The coupling of these functions is described by quantum interference parameters $\alpha_{kj}$ (21). As a result, virtual processes involving absorption and emission of photons at combination frequencies $|\Omega_{kj}|$ (21) occur when $\alpha_{kj} \gtrsim 1$ for a given number of absorbed and emitted photons (definite values of $n_1, \dots, n_N$). If the interference parameters are small ($\alpha_{kj} \ll 1$), then we can neglect the influence of correlated processes ($r_{12} = \ldots = r_{N-1,N} = 0$), and the functions $I_{n_1 \ldots n_j \ldots n_N}$ defined by (17) and (18) can be represented as products of Bessel functions describing independent emission and absorption of photons from the relevant waves:

$$I_{n_1 \ldots n_N}(\gamma_1, \ldots, \gamma_N; 0, \ldots, 0) = \exp\left(-i \sum_{j=1}^{N} \delta_j \chi_j n_j\right) \prod_{k=1}^{N} J_{n_k}(\gamma_k). \quad (22)$$

In what follows, we assume that the frequencies of the waves (e.g., $\omega_k$ and $\omega_j$) are not close to each other (otherwise, expressions (6) and (7) for the scattering amplitude are reduced to the formulas for the amplitude of electron–nucleus scattering in the field of ($N - 1$) waves [8]) and satisfy the conditions

$$\frac{|\omega_j - \omega_k|}{\omega_k} \gtrsim 1, \quad \omega_k \ll \begin{cases} m v_i^2/2, & v_i \ll 1 \\ m, & E_i \gtrsim m. \end{cases} \quad (23)$$

## 3. THE PROBABILITY OF ELECTRON–NUCLEUS SBEA

Using expressions (6) and (7) for the scattering amplitude, we can represent the probability of scattering of nonpolarized electrons by a nucleus in the presence of $N$ circularly polarized electromagnetic waves per unit time within an infinitely small volume of finite momenta $d^3p_f$ in the conventional form [10]

$$dW_{fi} = \left(\prod_{j=1}^{N} \sum_{n_j=-\infty}^{\infty}\right) dW_{fi}^{(n_1 \ldots n_N)}, \quad (24)$$

where the partial probability is given by

$$dW_{fi}^{(n_1 \ldots n_N)} = \frac{2(Ze^2)^2}{\tilde{E}_i \tilde{E}_f} H_{fi}^{(n_1 \ldots n_N)} \times \delta\left(\tilde{E}_f - \tilde{E}_i + \sum_{k=1}^{N} n_k \omega_k\right) \frac{d^3 p_f}{\mathbf{q}^4}. \quad (25)$$

Here, we introduced the notation

$$\begin{aligned} H_{fi}^{(n_1 \ldots n_N)} &= (m^2 + E_i E_f + \mathbf{p}_i \mathbf{p}_f)|I_{n_1 \ldots n_N}|^2 \\ &+ \frac{m^4}{2\kappa_i \kappa_f}|B_{n_1 \ldots n_N}|^2 + \frac{1}{4} f_1 m^2 |D_{n_1 \ldots n_N}|^2 \\ &+ 2m^2 \mathrm{Re}\Big\{ f_2 I^*_{n_1 \ldots n_N} B_{n_1 \ldots n_N} - B^*_{n_1 \ldots n_N}(\mathbf{f}_3 \mathbf{D}_{n_1 \ldots n_N}) \\ &+ \frac{1}{4\kappa_i \kappa_f}(\mathbf{p}_f \mathbf{D}_{n_1 \ldots n_N})(\mathbf{p}_i \mathbf{D}^*_{n_1 \ldots n_N}) - \frac{1}{4} I^*_{n_1 \ldots n_N}(\mathbf{f}_4 \mathbf{D}_{n_1 \ldots n_N}) \Big\}, \end{aligned} \quad (26)$$

where the functions $f_1$, $f_2$, $\mathbf{f}_3$, and $\mathbf{f}_4$ are written as

$$\begin{aligned} f_1 &= \frac{m^2 - E_i E_f + \mathbf{p}_i \mathbf{p}_f}{\kappa_i \kappa_f} + \frac{E_f \kappa_i - E_i \kappa_f}{\kappa_i}\left(\frac{1}{\kappa_f} - \frac{1}{\kappa_i}\right) \\ &- \frac{(\kappa_f - \kappa_i)^2 (E_i + \mathbf{n}\mathbf{p}_i)}{2\kappa_f \kappa_i^2}, \\ f_2 &= \frac{m^2 - E_i E_f + \mathbf{p}_i \mathbf{p}_f + E_f \kappa_i + E_i \kappa_f}{2\kappa_i \kappa_f}, \\ \mathbf{f}_3 &= \frac{m(\mathbf{p}_i + \mathbf{p}_f)}{4\kappa_i \kappa_f}, \\ \mathbf{f}_4 &= \frac{1}{m}\left[\frac{2E_i}{\kappa_i} + \left(\frac{1}{\kappa_f} - \frac{1}{\kappa_i}\right)(E_i + \mathbf{n}\mathbf{p}_i)\right]\mathbf{p}_f \\ &+ \frac{1}{m}\left[1 + \frac{2E_f - \kappa_f}{\kappa_i}\right]\mathbf{p}_i. \end{aligned} \quad (27)$$

Expressions (24)–(27) for the probability are valid for arbitrary intensities and frequencies of the waves and electron velocities $v_{i,f} \gg Z/137$. One can easily verify that, as ($N - 2$) waves are switched off (e.g., $F_1 = \ldots = F_{N-2} = 0$), the probability defined by (24)–(27) is reduced to the probability of electron–nucleus scattering in the field of two waves [8, 9]. If ($N - 1$) waves are switched off, then we arrive at the expressions for the probability of scattering in the field of a monochromatic plane wave [2]. Finally, if all $N$ waves are switched off, we obtain an expression for the conventional Mott probability of electron–nucleus scattering [10].

Note that, due to the complexity of energy-conservation relations, we cannot directly integrate the partial probability (25) in the energies of an electron in the final state for arbitrary wave intensities (a comprehensive analysis of the relevant equation was performed by Yakovlev [11]). However, we can easily integrate in the quasienergy of an electron in the final state using the relationship

$$d^3 p_f = Y d^3 \tilde{p}_f = Y \tilde{E}_f \sqrt{\tilde{E}_f^2 - m_*^2}\, d\tilde{E}_f d\tilde{\Omega}. \quad (28)$$

Here, $d\tilde{\Omega}$ is an infinitely small solid angle of the quasimomentum, $m_*$ is the effective mass of an electron in the field of $N$ waves, and $Y$ is the Jacobian of the transformation,

$$Y^{-1} = 1 + \frac{m^2}{2\kappa_f^2} \sum_{k=1}^{N} \eta_k^2, \quad m_* = m\sqrt{1 + \sum_{k=1}^{N} \eta_k^2}. \quad (29)$$

However, such an integration brings us into a nonphysical system of coordinates, where it is rather difficult to interpret experimental results.

## 4. ELECTRON–NUCLEUS SBEA IN THE BUNKIN–FEDOROV RANGE

Following studies [8, 9], we will call the kinematic region where quantum parameters $\gamma_k$ (20) are the main multiphoton parameters the Bunkin–Fedorov range. In this range, an electron is scattered in such a manner that the angle $\psi$ between the vector $\mathbf{g}_{fi}$ (20) and the direction of propagation of all $N$ waves is on the order of unity or, at least, is not too small (for $\gamma'_k \gg 1$); i.e.,

$$\psi \gtrsim \frac{1}{\gamma'_k}, \quad |\pi - \psi| \gtrsim \frac{1}{\gamma'_k},$$
$$\gamma'_k = \eta_k \frac{m|\mathbf{g}_{fi}|}{\omega_k}, \quad k = 1, \ldots, N. \tag{30}$$

Here, we assume that $\gamma'_k \gtrsim 1$ (otherwise, the intensities of the light waves are low, and we can apply perturbation theory with respect to the external field). As can be seen from (30), the Bunkin–Fedorov range corresponds to a rather broad kinematic region of scattering. In this region, the number of photons absorbed or emitted by an electron for each wave in the process of electron–nucleus scattering is determined by the corresponding quantum parameter $\gamma_k$ (20). Therefore, we mainly deal with processes where $|n_k| \lesssim \gamma_k$ (for $|n_k| \gg \gamma_k$, Bessel functions are very small). Consequently, the fraction of energy spent or absorbed by an electron due to emission or absorption of photons from each wave is subject to the following restriction:

$$\frac{n_k\omega_k}{E_i} \lesssim \xi_k,$$
$$\xi_k = \eta_k \frac{m}{|\mathbf{p}_i|} = \begin{cases} \eta_k/v_i, \; v_i \ll 1 \\ \eta_k(m/E_i), \; E_i \gg m \end{cases}, \quad k = 1, \ldots, N. \tag{31}$$

Obviously, the physical meaning of parameter $\xi_k$ is the ratio of the work done by the field within the spatial scale equal to the wavelength to the energy of the electron (in the nonrelativistic case, this is the ratio of the electron oscillatory velocity in the wave field to the velocity of translational motion). Note also that, within the Bunkin–Fedorov range, quantum interference parameters satisfy the relation

$$\alpha_{kj} \sim \gamma'_k \xi_j v_i. \tag{32}$$

Let us consider electron–nucleus SBEA in the range of moderately strong fields, when $\xi_k \ll 1$ [7, 8, 12, 13]. Depending on the electron energy, the intensities of the waves within this range are subject to the following restriction:

$$\eta_k \ll \begin{cases} v_i, \quad v_i \ll 1 \\ 1, \quad E_i \sim m \\ E_i/m, \quad E_i \gg m \end{cases} \tag{33}$$

By virtue of (33), we should keep only the first term in expression (26). Four-quasimomenta meet the relation $\tilde{p}_{i,f} \approx p_{i,f}$ [see (9)], and energy conservation is expressed as $E_f \cong E_i$ (except for the case of ultrarelativistic electrons in the initial or final state that move within a narrow cone of angles along the direction of wave propagation). Taking this circumstance into account, performing integration in energies of the electron in the final state, and dividing the probability (25) by the flow density of electrons incident on the nucleus, we derive the following ultimate expression for the partial cross section of electron–nucleus SBEA:

$$\frac{d\sigma^{(n_1\ldots n_N)}}{d\Omega} = |I_{n_1\ldots n_N}|^2 \frac{d\sigma_{\rm Mott}}{d\Omega}, \tag{34}$$

where $d\sigma_{\rm Mott}$ is the conventional cross section of electron–nucleus scattering in the absence of external fields (the Mott cross section) [10] and functions $I_{n_1\ldots n_N}$ are defined by expressions (17)–(21), where we should set $E_f = E_i$. As can be seen from (34), the cross section of multiphoton electron–nucleus SBEA in the range of intensities specified by (33) is factorized as a product of the probability to emit or to absorb photons from the relevant waves, which is determined by the modulus of functions $I_{n_1\ldots n_N}$ squared, by the cross section of elastic electron–nucleus scattering. Since quantum interference parameters $\alpha_{kj}$ can be estimated in their order of magnitude with the use of (32), provided that $\gamma_k \sim 1$, we have $\alpha_{kj} \ll 1$ in the range of intensities specified by (33). Hence, functions $I_{n_1\ldots n_N}$ are reduced to (22), and, consequently, we have

$$|I_{n_1\ldots n_N}|^2 = J^2_{n_1}(\gamma_1)J^2_{n_2}(\gamma_2)\ldots J^2_{n_N}(\gamma_N) \tag{35}$$

in (34).

If the quantum parameters satisfy the inequality $\gamma_k \gg 1$ [and condition (33) is met], then we may have $\alpha_{kj} \gtrsim 1$. Using the asymptotic representation of Bessel functions [14], we can readily show that, in this case, the modulus of functions $I_{n_1\ldots n_N}$ squared can be estimated in its order of magnitude as

$$|I_{n_1\ldots n_N}|^2 \sim (\gamma_1\ldots\gamma_N)^{-1} \ll 1; \tag{36}$$

i.e., the partial cross sections (34) are small as compared with the Mott cross section.

Since the arguments of the functions $I_{n_1\ldots n_N}$ defined by (17) and (18) are independent of summation indices within the range specified by (33), we can easily perform summation in the partial cross section (34) over all possible processes involving emission or absorption of photons from the relevant waves:

$$\frac{d\sigma}{d\Omega} = \frac{d\sigma_{\rm Mott}}{d\Omega}\left[\prod_{k=1}^{N}\sum_{n_k=-\infty}^{\infty}|I_{n_1\ldots n_N}|^2\right] = \frac{d\sigma_{\rm Mott}}{d\Omega}. \tag{37}$$

Thus, performing appropriate summation, we find that all essentially quantum contributions to the intensities (33) compensate each other.

Note that in the limiting case of nonrelativistic electron velocities, the upper inequality in (33) is generally insufficient to ensure the applicability of the dipole approximation in the interaction of an electron with the electric fields of the waves. The necessary condition for the validity of such an approximation is that interference parameters $\alpha_{kj}$ (21) (these parameters are equal to zero in the dipole approximation) should be small. This condition is satisfied when the products of wave intensities meet the following inequality:

$$\eta_k\eta_j \ll \frac{\omega_{k(j)}}{m v_i}. \quad (38)$$

This condition essentially involves the frequencies of the waves and, generally, may be more stringent than the upper inequality in (33). Specifically, for optical frequencies ($\omega_k \sim 10^{15}$ s$^{-1}$) and velocities $v_i = 10^{-1}$, we can use (33) and (38) to obtain $\eta_k \ll 10^{-1}$ and $\eta_k\eta_j \ll 10^{-5}$, respectively. In other words, the applicability of the dipole approximation (38) requires wave intensities much lower than those required for the fulfillment of (33). However, we can always choose the frequency range in such a way that condition (38) is *a fortiori* satisfied within the framework of (33), and the dipole approximation is applicable. In this case, the first factor in the partial cross section (34) is reduced to (35).

Now, let us consider the scattering of a nonrelativistic electron in the range of strong fields, when $\xi_k \gtrsim 1$. By virtue of this condition, oscillatory velocities of electrons in wave fields are on the order of or higher than the velocity of electron translational motion,

$$\eta_k \gtrsim v_i \ll 1. \quad (39)$$

We will also assume that

$$\eta_k \ll 1, \quad k = 1, 2, \ldots, N. \quad (40)$$

Taking (39) and (40) into account, we can write the law of energy conservation as

$$\frac{\mathbf{p}_f^2}{2m} - \frac{\mathbf{p}_i^2}{2m} + \frac{1}{2}\left(\sum_{k=1}^{N}\eta_k^2\right)\mathbf{n}(\mathbf{p}_f - \mathbf{p}_i) + \sum_{k=1}^{N} n_k\omega_k = 0. \quad (41)$$

Note that the third term in (41) is of the same order of magnitude as other terms when wave intensities meet the condition

$$\eta_k^2 \gtrsim v_i, \quad (42)$$

i.e., when electron oscillatory velocities in the wave field are much higher than the translational velocity ($\eta_k \gg v_i$). The law of energy conservation (41) is a quadratic equation with respect to the velocity of an electron in the final state. Performing simple calculations, we can find possible values of electron velocities in the final state:

$$v_f = \begin{cases} v_1 = -b + \sqrt{b^2 + a}, & a > 0 \\ v_\pm = |b| \pm \sqrt{b^2 + a}, & a < 0, \quad \theta_f > \pi/2. \end{cases} \quad (43)$$

Here, we introduced the following notations:

$$a = v_i^2 + \left(\sum_{k=1}^{N}\eta_k^2\right) v_i \cos\theta_i - \frac{2}{m}\sum_{k=1}^{N} n_k\omega_k, \quad (44)$$

$$b = \frac{1}{2}\left(\sum_{k=1}^{N}\eta_k^2\right)\cos\theta_f, \quad \theta_{i,f} = \angle(\mathbf{n}, \mathbf{p}_{i,f}). \quad (45)$$

Using these expressions and performing integration in the velocities of an electron in the final state in the partial probability described by (25) and (26), we derive the following expression for the cross section of scattering:

$$\frac{d\sigma_{fi}^{(n_1\ldots n_N)}}{d\Omega} = 4Z^2 r_e^2 \frac{v_f^2}{v_i(v_f + b)(\mathbf{v}_f - \mathbf{v}_i)^4}\left|I_{n_1\ldots n_N}\right|^2. \quad (46)$$

Here, $r_e$ is the classical electron radius; quantum interference parameters in the functions $I_{n_1\ldots n_N}$ defined by (17) and (18) are written as

$$\alpha_{kj} = \eta_k\eta_j\frac{(\mathbf{n}\mathbf{g}_{fi})}{\Omega_{kj}}, \quad \mathbf{g}_{fi} = \mathbf{v}_f - \mathbf{v}_i; \quad (47)$$

and $\gamma_k$ and $\chi_k$ are given by expressions (20) and (19) with a vector $\mathbf{g}_{fi}$ (47). In the partial cross section (46), we have $v_f = v_1$ (43). If equation (41) has two roots, then the sought-for cross section can be represented as a sum of two cross sections (46) with $v_f = v_+$ and $v_f = v_-$.

As can be seen from (47), for intensities that satisfy condition (42), interference parameters meet the relations $\alpha_{kj} \sim m v_i^2/\omega_{k(j)} \gg 1$ and $\gamma_k \gg \alpha_{kj}$ in the range of optical frequencies. By virtue of these relations, multiquantum processes with absorption of $|n_k| \sim \gamma_k$ ($k = 1, 2, \ldots, N$) photons from the waves are dominant within the considered range of wave intensities and frequencies (emission of such a large number of photons is forbidden by the law of energy conservation).

Next, let us consider the case when wave intensities meet the inequality opposite of (42), i.e.,

$$\eta_k^2 \ll v_i, \quad (48)$$

but condition (39) is satisfied. Provided that (48) is true, we can neglect the third term in the law of energy conservation (41). Then, we can derive the expression for the partial cross section by setting $b = 0$ in (46) and neglecting the second term in the coefficient $a$ defined by (44):

$$\frac{d\sigma_{fi}^{(n_1\ldots n_N)}}{d\Omega} = 4Z^2 r_e^2 \frac{v_f}{v_i(\mathbf{v}_f - \mathbf{v}_i)^4}|I_{n_1\ldots n_N}|^2,$$
$$v_f = v_i\sqrt{1 - \frac{2}{mv_i^2}\sum_{k=1}^{N} n_k\omega_k}. \tag{49}$$

Note that, generally, the interference parameters are not small in the case under study: $mv_i^3/\omega_{k(j)} \lesssim \alpha_{kj} \ll mv_i^2/\omega_{k(j)} \gg 1$. We have $\alpha_{kj} \ll 1$ and the functions $|I_{n_1\ldots n_N}|^2$ in (49) are reduced to (35) only for intensities that satisfy the condition of dipole approximation (38).

## 5. ELECTRON–NUCLEUS SBEA IN THE INTERFERENCE RANGE

Let us consider the scattering of an electron by a nucleus in the field of $N$ waves under conditions when the angle $\psi$ between the wave vector and the vector $\mathbf{g}_{fi}$ [see (20)] is so small that Bunkin–Fedorov quantum multiphoton parameters are small for arbitrary intensities of the waves; i.e., scattering occurs in the kinematic range specified by conditions opposite of (30):

$$\psi \ll \frac{1}{\gamma_k'}, \quad |\pi - \psi| \ll \frac{1}{\gamma_k'}, \quad k = 1, 2, \ldots, N, \tag{50}$$

where $\mathbf{g}_{fi}^2 = (\mathbf{n}\mathbf{g}_{fi})^2$. In what follows, this kinematic range will be referred to as the interference range. Note that, by summing all $\lambda_k$ ($k = 1, 2, \ldots, N$) [see (19)], we derive an equality

$$\sum_{k=1}^{N}\delta_k\lambda_k = -\sum_{j=1}^{N}\delta_j n_j. \tag{51}$$

In the interference range, we can set $\gamma_1 = \ldots = \gamma_N = 0$ and $\chi_1 = \ldots = \chi_N = 0$ [see (19) and (20)]. With allowance for these equalities, we find from (17)–(19) that $\lambda_k = 0$. Therefore, employing (51), we derive an equation that couples the numbers of photons in all the waves:

$$\sum_{k=1}^{N}\delta_k n_k = 0. \tag{52}$$

Hence, the photon numbers $n_k$ in the interference range are correlated in such a manner that the total number of photons (with allowance for polarization $\delta_k$) emitted or absorbed by an electron is equal to zero. As demonstrated below, the ambiguity in the choice of the number of the wave involved in emission (absorption) of photons correlated with other waves has no influence on the probability of the process. It is due to relation (52) that emission and absorption of photons at combination frequencies $|\Omega_{kN}|$ occur in the interference range. Note that quantum parameters $\alpha_{kN}$ (21) play the role of multiphoton parameters in this case. Therefore, it would be natural to analyze phenomena that occur in the interference range in terms of a set of $(N-1)$ numbers of photons at combination frequencies $|\Omega_{kN}|$ rather than in terms of numbers $n_1, \ldots, n_N$ of emitted and absorbed photons at frequencies $\omega_1, \ldots, \omega_N$. Thus, we introduce photon numbers $l_1, \ldots, l_{N-1}$ defined as

$$\begin{cases} 2l_0 = \delta_1 n_1 + \ldots + \delta_N n_N \\ -2l_1 = -\delta_1 n_1 + \delta_2 n_2 + \ldots + \delta_N n_N \\ -2l_2 = \delta_1 n_1 - \delta_2 n_2 + \ldots + \delta_N n_N \\ \ldots\ldots\ldots\ldots\ldots\ldots\ldots\ldots\ldots \\ -2l_{N-1} = \delta_1 n_1 + \ldots + \delta_{N-2}n_{N-2} - \delta_{N-1}n_{N-1} + \delta_N n_N. \end{cases} \tag{53}$$

By virtue of (52) and (53), we have $l_0 = 0$ and

$$n_k = \delta_k l_k \quad (k = 1, 2, \ldots, N-1),$$
$$n_N = -\delta_N\sum_{k=1}^{N-1} l_k. \tag{54}$$

With allowance for these relations, the law of energy conservation is written as [see the argument of the $\delta$-function in (25)]

$$\tilde{E}_f - \tilde{E}_i + \sum_{k=1}^{N-1} l_k\Omega_{kN} = 0. \tag{55}$$

Then, we can eliminate $(N-1)$ sums in $r_{jN}$ from expression (17). Thus, we have only

$$N_3 = \frac{1}{2}(N-1)(N-2) \tag{56}$$

sums, and the functions $I_{n_1\ldots n_N}$ defined by (17)–(19) are reduced to functions $J_{l_1\ldots l_{N-1}}$ that depend only on $(N-1)$ interference quantum parameters and that can be represented as

$$J_{l_1\ldots l_{N-1}}(\alpha_{1N}, \ldots, \alpha_{N-1,N})$$
$$\equiv I_{l_1\ldots l_{N-1}}(0, \ldots, 0; \alpha_{1N}, \ldots, \alpha_{N-1,N})$$
$$= \left\{\prod_{j=1}^{N-2}\prod_{k=j+1}^{N-1}\sum_{r_{jk}=-\infty}^{\infty} J_{r_{jk}}(\alpha_{jk})\right\}\left[\prod_{s=1}^{N-1} J_{r_{sN}}(\alpha_{sN})\right]. \tag{57}$$

Here, integer orders $r_{sN}$ of Bessel functions are determined from relations

$$r_{sN} = l_s + \sum_{k=1}^{s-1} r_{ks} - \sum_{k=s+1}^{N-1} r_{sk}, \tag{58}$$

and parameters $\alpha_{jk}$ can be expressed in terms of products of parameters $\alpha_{sN}$:

$$\alpha_{jk} = \eta_j\eta_k\frac{\alpha_{jN}\alpha_{kN}}{\eta_N(\eta_j\alpha_{kN} - \eta_k\alpha_{jN})}. \tag{59}$$

As can be seen from expressions (55)–(59) derived above, the scattering of an electron in the interference range is characterized by a strong correlation between the numbers of photons emitted or absorbed by the electron (52). Due to this circumstance, multiphoton electron–nucleus SBEA in this range is governed by quantum interference parameters, as opposed to the Bunkin–Fedorov range, where quantum parameters $\gamma_k$ (20) play the role of multiphoton parameters. In the interference range, we have $(N-1)$ independent multiphoton parameters $\alpha_{sN}$, which characterize emission and absorption of photons at combination frequencies $|\Omega_{kN}|$. Note that all these parameters are coupled with each other through parameters $\alpha_{jk}$ (59).

Note that, in the case of two waves, the functions $J_{l_1 \ldots l_{N-1}}$ defined by (57) can be expressed in terms of a single Bessel function $J_{l_1} = J_{l_1}(\alpha_{12})$; i.e., we deal with correlated emission and absorption of equal numbers of photons from both waves (see (52), (53) and [8]). In the case of three waves, we have

$$J_{l_1 l_2}(\alpha_{13}, \alpha_{23}) = \sum_{r=-\infty}^{\infty} J_r(\alpha_{12}) J_{l_1 - r}(\alpha_{13}) J_{l_2 + r}(\alpha_{23}), \quad (60)$$

where

$$\alpha_{12} = \eta_1 \eta_2 \frac{\alpha_{13}\alpha_{23}}{\eta_3(\eta_1\alpha_{23} - \eta_2\alpha_{13})}, \quad (61)$$

and the law of energy conservation (55) is written as

$$\tilde{E}_f - \tilde{E}_i + l_1\Omega_{13} + l_2\Omega_{23} = 0. \quad (62)$$

We emphasize that the ambiguity in the notation of the wave number has no influence on the probabilities of the processes under study. For simplicity, we will prove this statement for the case of three waves. Using the replacement $\omega_1 \longleftrightarrow \omega_3$, we can rewrite the functions $J_{l_1 l_2}$ (60) and the law of energy conservation (62) as

$$J_{l_1 l_2} = \sum_{r=-\infty}^{\infty} J_r(\alpha_{32}) J_{l_1 - r}(\alpha_{31}) J_{l_2 + r}(\alpha_{21}), \quad (63)$$

$$\tilde{E}_f - \tilde{E}_i + l_1\Omega_{31} + l_2\Omega_{21} = 0. \quad (64)$$

As can be seen from (62) and (64), wave renumbering changes the form of the energy-conservation law along with the form of the functions $J_{l_1 l_2}$. Let us rename the summation index in (63), $l_2 + r = -r'$, and employ the replacement $l_1 + l_2 \longrightarrow -l_1'$. Then, with allowance for the properties of Bessel functions $[J_{-r'}(\alpha_{21}) = J_{r'}(\alpha_{12})]$, we can readily show that (63) and (64) are reduced to initial expressions (60) and (62).

As it follows from conditions (50), the interference range is characterized by the scattering of an electron in the plane formed by the initial momentum of the electron and the wave vector. The azimuthal angles of an electron in the initial and final states are equal to each other, and the polar angles and the velocities are related by the following expression (see [8]):

$$v_f = \frac{a_i}{\sin\theta_f + a_i\cos\theta_f}, \quad a_i = \frac{v_i \sin\theta_i}{1 - v_i\cos\theta_i}. \quad (65)$$

With allowance for (65), the law of energy conservation (55) can be considerably simplified and can be written in the form of a quadratic equation (instead of a fourth-order equation) with respect to the energy of the emerging electron:

$$E_f^2 - 2(b_i E_i)E_f + b_0 m^2(1 + a_i\cot\theta_f) = 0. \quad (66)$$

Here, we introduced the following notations:

$$2b_i = 1 + b_0\frac{m^2}{\kappa_i E_i} - \frac{1}{E_i}\sum_{k=1}^{N-1} l_k\Omega_{kN}, \qquad b_0 = \frac{1}{2}\sum_{k=1}^{N}\eta_k^2. \quad (67)$$

By virtue of (32), we can write

$$\gamma_k' \sim \alpha_{kj}\frac{E_i}{m\eta_j} = \begin{cases} \alpha_{kj}/\eta_j, & v_i \ll 1 \\ \alpha_{kj}/\xi_j, & E_i \gg m. \end{cases} \quad (68)$$

Hence, for nonrelativistic and relativistic electrons (when $\eta_j \ll 1$), as well as for ultrarelativistic electron energies (when $\xi_j \ll 1$), we have

$$\gamma_k' \gg \alpha_{kj} \gtrsim 1 \quad (69)$$

In this case, we can easily demonstrate (see [13]) that

$$\frac{|I_{n_1 \ldots n_N}|^2}{|J_{l_1 \ldots l_{N-1}}|^2} \sim \begin{cases} (\gamma_1 \ldots \gamma_N)^{-1} \ll 1, & l_k \ll \gamma_k \\ (\gamma_1 \ldots \gamma_N)^{-2/3} \ll 1, & l_k \sim \gamma_k, \end{cases} \quad k = 1, \ldots, N. \quad (70)$$

Comparison of the probabilities of multiphoton SBEA in the kinematic ranges of electron scattering specified by (30) and (50) shows that the ratio of these probabilities is of the same order of magnitude as the ratio of functions (70). Consequently, if conditions (69) are satisfied, the probability of multiphoton electron–nucleus SBEA in the interference range is considerably higher than the probability of multiphoton electron–nucleus SBEA in the Bunkin–Fedorov range. Note also that, in the interference range, the classical interference parameter

$$\xi_{jk} = \xi_j\xi_k(|\mathbf{p}_i|/E_i) \quad (71)$$

plays the role of the classical parameter $\xi_k$ (31) in the Bunkin–Fedorov range.

Let us sequentially consider electron–nucleus SBEA in the interference range specified by (50) for relativistic, nonrelativistic, and ultrarelativistic electron energies.

Suppose that an electron has a relativistic energy ($E_{i,f} \sim m$) and the intensities of the waves are limited by condition (40). Then, using the law of energy conservation (66), we find that $E_f = E_i$, and formulas (65) yield the following expressions for the angle of electron scattering:

$$\tan\frac{\theta}{2} = \begin{cases} (\cos\theta_i - v_i)/\sin\theta_i, & \theta_i < \pi/2 \\ (|\cos\theta_i| + v_i)/\sin\theta_i, & \theta_i > \pi/2. \end{cases} \quad (72)$$

Employing these results, we can readily derive the cross section of scattering from (24) and (25):

$$\frac{d\sigma}{d\Omega} = \left(\prod_{k=1}^{N-1}\sum_{l_k=-\infty}^{\infty}\right)\frac{d\sigma^{(l_1\ldots l_{N-1})}}{d\Omega}. \quad (73)$$

Here, the partial cross section of scattering with emission or absorption of $l_1, \ldots, l_{N-1}$ photons at combination frequencies $|\Omega_{1N}|, \ldots, |\Omega_{N-1,N}|$ is written as

$$\frac{d\sigma^{(l_1\ldots l_{N-1})}}{d\Omega} = |J_{l_1\ldots l_{N-1}}|^2\frac{d\sigma_{\text{Mott}}}{d\Omega}, \quad (74)$$

where the functions $J_{l_1\ldots l_{N-1}}$ are given by (57). Note that the cross section (74) is considerably greater than the corresponding cross section in the Bunkin–Fedorov range (34) for intensities (69). Summing partial cross sections (74) over all processes involving emission and absorption of photons from the relevant waves, we find that all essentially quantum contributions characterized by quantum interference parameters compensate each other, and the total cross section coincides with the Mott cross section.

Next, let us consider electrons with nonrelativistic energies ($v_{i,f} \ll 1$), assuming that wave intensities satisfy (40). Performing simple transformations, we can reduce the law of energy conservation (66) to the relation

$$\frac{\mathbf{p}_f^2}{2m} - \frac{\mathbf{p}_i^2}{2m} + \sum_{k=1}^{N-1} l_k\Omega_{kN} = 0. \quad (75)$$

Taking this formula into account and using (65), we can find the polar angle of the emerging electron:

$$\sin\theta_f = \rho_{l_1\ldots l_{N-1}}^{-1}\sin\theta_i, \quad (76)$$

where

$$\rho_{l_1\ldots l_{N-1}} = \frac{v_f}{v_i} = \sqrt{1 - \frac{2}{mv_i^2}\sum_{k=1}^{N-1} l_k\Omega_{kN}}. \quad (77)$$

Hence, the scattering angle generally depends on the number of absorbed and emitted photons at combination frequencies, and the polar angles of the incident electron may satisfy (if $\rho_{l_1\ldots l_{N-1}} < 1$) the following inequality:

$$\sin\theta_i < \rho_{l_1\ldots l_{N-1}}. \quad (78)$$

It can be easily seen that, for sufficiently strong fields, when $\eta_j\eta_k \gg v_i$ [but conditions (40) are satisfied], the second term in the radicand in (77) can be large; i.e., multiphoton absorption of photons at combination frequencies may dominate. As a result, the velocity of the emerging electrons may considerably exceed the initial electron velocity ($\rho_{l_1\ldots l_{N-1}} \gg 1$). Then, by virtue of (76), we find that, regardless of the initial polar angle $\theta_i$ of the incident electron, the emerging electron is scattered either along the wave vector or in the direction opposite of the wave vector ($\theta_f \approx 0, \pi$). Suppose now that the product of wave intensities meets the condition

$$\eta_j\eta_k \sim v_i. \quad (79)$$

Then, the scattering cross section for nonrelativistic electrons scattered by angles (76) in the interference range (50) can be derived from (25):

$$\frac{d\sigma^{(l_1\ldots l_{N-1})}}{d\Omega} = 4Z^2r_e^2\frac{m^4}{(\mathbf{p}_f - \mathbf{p}_i)^4}\rho_{l_1\ldots l_{N-1}}|J_{l_1\ldots l_{N-1}}|^2, \quad (80)$$

where the interference quantum parameters in functions $J_{l_1\ldots l_{N-1}}$ are given by

$$\alpha_{kN} = \eta_k\eta_N\frac{mv_i}{\Omega_{kN}}(\rho_{l_1\ldots l_{N-1}}\cos\theta_f - \cos\theta_i). \quad (81)$$

If the intensities of the waves meet the conditions

$$\eta_j\eta_k \ll v_i, \quad (82)$$

we can neglect the number of emitted (absorbed) photons of the waves. In other words, under these conditions, we have $v_f \approx v_i$, and an electron is scattered by an angle $\theta_f = \pi - \theta_i$ or

$$\theta = \begin{cases} \pi - 2\theta_i, & \theta_i < \pi/2 \\ 2\theta_i - \pi, & \theta_i > \pi/2. \end{cases} \quad (83)$$

Therefore, expressions (80) and (81) for the cross section and interference parameters can be rewritten as

$$\frac{d\sigma^{(l_1\ldots l_{N-1})}}{d\Omega} = \frac{Z^2r_e^2}{4v_i^4\cos^4\theta_i}|J_{l_1\ldots l_{N-1}}|^2, \quad (84)$$

$$\alpha_{kN} = -2\eta_k\eta_N\frac{mv_i\cos\theta_i}{\Omega_{kN}}. \quad (85)$$

Summation of the cross section (84) over all emission and absorption processes involving photons at combination frequencies yields the Mott cross section. Note that, if the wave intensities are chosen in such a manner that $\alpha_{kN} \gtrsim 1$, the partial cross sections (80) and (84) are considerably greater than the corresponding cross sections of scattering in the Bunkin–Fedorov range. Specifically, for optical frequencies of the waves and velocities $v_i = 10^{-1}$, such a relation between the cross sections of scattering holds within the range of wave intensities $10^{-5} < \eta_k\eta_N < 10^{-1}$, i.e., for intensities that can be achieved with modern lasers.

Now, let us consider the case when electrons have ultrarelativistic energies ($|\mathbf{p}_{i,f}| \approx E_{f,i} \gg m$) and the wave intensities meet the conditions $\xi_k \ll 1$, i.e.,

$$\eta_k \ll E_i/m, \quad k = 1, 2, \ldots, N. \tag{86}$$

In this case, formula (65) yields the following expression for the interference range:

$$\frac{v_f \sin\theta_f}{(1 - v_f) + 2\sin^2(\theta_f/2)} = \frac{v_i \sin\theta_i}{(1 - v_i) + 2\sin^2(\theta_i/2)}. \tag{87}$$

We will analyze the scattering of an electron by large angles and exclude the case when $\theta_i \sim 1$ and $\theta_f \sim 1$. Indeed, by virtue of (66), the electron energy before scattering is equal to the electron energy after scattering [$E_i = E_f$, see the argument of the δ-function in (25)], and it is easy to verify that relation (87) is reduced to the equality $\cot(\theta_f/2) = \cot(\theta_i/2)$. This equality can be satisfied only when an electron is scattered by zero angle. Therefore, in the studied case of large-angle scattering, relation (87) is met only in two cases: when $\theta_i \ll 1$ and $\theta_f \sim 1$ or $\theta_i \sim 1$ and $\theta_f \ll 1$. In the latter case, we find that, for arbitrary initial angles $\theta_i$, the direction of electron scattering is close to the direction of propagation of both waves:

$$\begin{aligned}\theta_f &= (1 - v_f)(v_i/v_f)\cot\frac{\theta_i}{2} \ll 1;\\ \theta &= \theta_i - \theta_f \approx \theta_i \sim 1.\end{aligned} \tag{88}$$

By virtue of (86) and (88), we can set $\tilde{E}_i = E_i$ and $\tilde{E}_f = E_f(1 + 2b_0)$ in the law of energy conservation (66), and the energy of the scattered electron is given by the following expressions (see also [9]):

$$\begin{aligned}E_f &= \rho_{l_1\ldots l_{N-1}} E_i;\\ \rho_{l_1\ldots l_{N-1}} &= (1 + 2b_0)^{-1}\left[1 - \frac{1}{E_i}\sum_{k=1}^{N-1} l_k\Omega_{kN}\right].\end{aligned} \tag{89}$$

Using these formulas, we can rewrite expression (88) for scattering angles as

$$\theta_f = (1 - v_i)\frac{\cot(\theta_i/2)}{\rho^2_{n_1\ldots n_{N-1}} - (1 - v_i)} \ll 1. \tag{90}$$

As can be seen from (88)–(90), $E_f \approx E_i$ ($\rho_{l_1\ldots l_{N-1}} \approx 1$) for intensities $\eta_k^2 \ll 1$. Consequently, scattering angles (90) are given by

$$\theta_f = (1 - v_i)\cot(\theta_i/2) \sim (1 - v_i) \tag{91}$$

and are independent of the number of emitted and absorbed photons in both waves. In the opposite limiting case of very strong fields ($\eta_k^2 \gg 1$), expressions (89) yield

$$\rho_{l_1\ldots l_{N-1}} \approx \frac{1}{2b_0}\left[1 - \frac{1}{E_i}\sum_{k=1}^{N-1} l_k\Omega_{kN}\right] \sim \eta_k^{-2} \ll 1. \tag{92}$$

Consequently, scattering angles (90) can be estimated in their order of magnitude as $\theta_f \sim \eta_k^4(1 - v_i) \gg (1 - v_i)$, and the energies of emerging electrons (89) are low as compared with the energies of incident electrons ($E_f \ll E_i$). Since we assume that electrons in the final state are also ultrarelativistic, relation (92) imposes a more stringent condition on the intensities of both waves than (86):

$$\eta_k^2 \ll E_i/m. \tag{93}$$

To find the cross section of scattering, we divide the probability (25) by the flow density of incident electrons. Taking (93) into account and performing integration in the energy of electrons in the final state, we derive the following ultimate expression for the partial differential cross section of scattering into an elementary solid angle $d\Omega$:

$$\frac{d\sigma^{(l_1\ldots l_{N-1})}}{d\Omega} = \frac{\rho^2_{l_1\ldots l_{N-1}}}{(1 + 2b_0)^2}\left|J_{l_1\ldots l_{N-1}}\right|^2\frac{d\sigma_{\text{Mott}}^{(E \gg m)}}{d\Omega}. \tag{94}$$

Here,

$$\frac{d\sigma_{\text{Mott}}^{(E \gg m)}}{d\Omega} = Z^2 r_e^2\left(\cot\frac{\theta}{2}/2\sin\frac{\theta}{2}\right)^2(m/E_i)^2 \tag{95}$$

is the ultrarelativistic limit of the Mott cross section [10]. Functions $J_{l_1\ldots l_{N-1}}$ are defined by expression (57), and the arguments of these functions are

$$\alpha_{kN} = \eta_k\eta_N\frac{E_i}{\Omega_{kN}}\rho_{l_1\ldots l_{N-1}}. \tag{96}$$

Hence, we find that, for $\eta_k \gtrsim 1$, quantum parameters meet the relations $\alpha_{kN} \gtrsim E_i/\omega_k \gg 1$, and, consequently, partial sections (94) are small as compared with the Mott cross section. Taking this circumstance into account, we will consider in greater detail the case when $\eta_k \ll 1$. In this regime, we have $E_f \approx E_i$, and the scattering angles $\theta_f$ and quantum interference parameters are given by (91) and (96) with $\rho_{l_1\ldots l_{N-1}} = 1$. Then, partial cross sections (94) are written as

$$\frac{d\sigma^{(l_1\ldots l_{N-1})}}{d\Omega} = \left|J_{l_1\ldots l_{N-1}}\right|^2\frac{d\sigma_{\text{Mott}}^{(E \gg m)}}{d\Omega}. \tag{97}$$

Note that the interference parameters meet the relation $\alpha_{kN} \geq 1$ when the product of wave intensities satisfies the condition

$$\eta_k\eta_N \geq \omega_k/E_i. \tag{98}$$

Note that, in the Bunkin–Fedorov range (30), quantum multiphoton parameters meet the conditions $\gamma_k \geq \eta_k(m/\omega_k) \geq m/\sqrt{E_i\omega_k}$. Therefore, provided that

$$m/\sqrt{E_i\omega_k} \gg 1, \qquad (99)$$

we have $\gamma_k \gg 1$, and relation (70) holds true; i.e., the cross section (97) of multiphoton electron–nucleus SBEA in the ultrarelativistic case is considerably greater than the relevant cross section of scattering for any other geometry. In other words, ultrarelativistic electrons are mainly scattered along the direction of propagation of $N$ waves (91).

Let us estimate field intensities and energies of electrons that satisfy conditions (98) and (99). In the range of optical frequencies, inequality (99) yields $\sqrt{E_i/m} \ll 10^2$–$10^3$. Therefore, using (98) with $E_i/m \sim 10^2$, we obtain field strengths $F_k \sim (10^8$–$10^6)$ V/cm, which can be readily achieved with modern lasers.

## 6. CONCLUSION

The performed investigation of SBEA that accompanies the scattering of an electron by a nucleus in the field of an arbitrary number $N$ of circularly polarized waves with arbitrary intensities and frequencies propagating in the same direction allows us to make the following conclusions:

(1) The probability of multiphoton SBEA involving emission and absorption of $n_1, \ldots, n_N$ photons of the waves is governed by the function $I_{n_1 \ldots n_N}$ defined by (17) and (18). Generally, this function depends on $N$ Bunkin–Fedorov quantum parameters $\gamma_k$ (20) and $N_1$ (1) quantum interference parameters $\alpha_{kj}$ (21).

(2) Depending on the intensities and frequencies of the waves, multiphoton SBEA may occur in substantially different kinematic ranges of electron scattering: the Bunkin–Fedorov range, where quantum parameters $\gamma_k$ serve as multiphoton parameters, and the interference range, where the interference of all the waves becomes clearly pronounced, and quantum parameters $\alpha_{kN}$ play the role of multiphoton parameters.

(3) For moderately strong fields [$\xi_k \ll 1$, $k = 1, \ldots, N$ (31)], the cross section of multiphoton electron–nucleus SBEA in the general relativistic case can be factorized in the Bunkin–Fedorov range (30) as a product of the probability to emit or to absorb a photon from the relevant wave, which is determined by the modulus of the functions $I_{n_1 \ldots n_N}$ squared, by the cross section of elastic electron–nucleus scattering (the Mott cross section). In the limiting case of nonrelativistic electron velocities, the applicability condition of dipole approximation (38) considerably depends on wave frequencies.

(4) In the interference range, the numbers of photons in all $N$ waves emitted and absorbed by an electron are strongly correlated with each other. Due to this correlation, the scattering of an electron by a given angle in the plane of the initial momentum and the wave vector is accompanied by the emission or absorption of photons at combination frequencies $|\Omega_{kN}|$ (21). For wave intensities and frequencies that satisfy the conditions $\alpha_{kN} \gtrsim 1$ and $\eta_k \ll 1$, the probability of such multiphoton SBEA is considerably higher than the SBEA probability in the Bunkin–Fedorov range. In the range of optical frequencies, conditions specified above correspond to the following restrictions on the field strengths:

$$(10^7\text{–}10^8)\ \text{V/cm} \lesssim F_k \ll (10^{10}\text{–}10^{11})\ \text{V/cm}$$

for electrons with relativistic energies and

$$(10^7\text{–}10^8)v_i\ \text{V/cm} \lesssim F_k \ll (10^{10}\text{–}10^{11})\ \text{V/cm}$$

for electrons with nonrelativistic energies.

## ACKNOWLEDGMENTS

We are grateful to Professor V.P. Krainov and Professor M.V. Fedorov for useful discussions and valuable remarks.